\documentclass{article}
\usepackage{color}
 \usepackage{amsbsy}
  \usepackage{bm}
 \usepackage{amsfonts}
\usepackage{amssymb}
\usepackage{amsmath}
\usepackage{graphics}
\usepackage{graphicx}

\begin{document}

\title{Hadronic matter under an external magnetic field: In-medium modification of the pion mass}
\author{R. M. Aguirre \\
Departamento de Fisica, Facultad de Ciencias Exactas, \\Universidad Nacional de La Plata, \\
and Instituto de Fisica La Plata, CONICET. Argentina.}

\date{}

\maketitle

\begin{abstract}
The covariant propagator of a fermion with intrinsic magnetic moment
interacting with a uniform external magnetic field is presented for
finite temperature and baryonic density. The case of a scalar boson
is also considered. The final expressions are given in terms of a
four-dimensional momentum representation. These results, which take
account of the full effect of the magnetic field, are used to
evaluate the modification of the pion mass at zero temperature as a
function of the density and the magnetic intensity. For this purpose
a self-consistent calculation, including one- and two-pion vertices,
is employed.

\noindent
\\

\end{abstract}

\newpage

\section{Introduction}

The dynamics of matter subject to strong magnetic fields has been
widely studied in the past \cite{LAI}, and it has received renewed
interest due to the analysis of different experimental situations \cite{MIRANSKY}.\\
In recent years the significative role played by the intrinsic
magnetic moments of the hadrons when the thermodynamical behavior of
dense nuclear matter under strong magnetic fields is analyzed has
been pointed out.
\cite{DONG1,DONG2,AB&V,RABHI,REZAEI}.\\
This conclusion can be made extensive, for instance, to the study
of matter created in heavy ion collisions, where very intense
magnetic fields have been predicted \cite{KHARZEEV,MO,SKOKOV}.
Experimental evidence of this fact is the preferential emission of
charged particles along the direction of the magnetic field for
noncentral heavy ion collisions, due to magnetic intensities $e\,
B \sim 10^2$
MeV$^2$\cite{KHARZEEV}. \\
In a different scenario, very dense hadronic matter under strong
magnetic fields could be found in certain compact stars, which have
generally been included within the magnetar model
\cite{DUNCAN,THOMPSON}. The sustained x-ray luminosity in the soft
(0.5-10 keV) or hard (50-200 keV) spectrum, as well as the bursting
activity of these objects are attributed to the dissipation and
decay of very strong magnetic fields. The intensity of these fields
has been estimated around $10^{15}$ G at the star surface, but could
reach much higher values in the dense interior of the star. The
availability of an increasing amount of precision data opens the
question on how well the current theoretical description of hadronic
matter can fit this empirical evidence.

A successful description of the dense hadronic environment has been
given by  a covariant model of the hadronic interaction known as
Quantum Hadro-Dynamics (QHD) \cite{SW}. It has been used to study
the structure of neutron stars and particularly to analyze hadronic
matter in the presence of an external magnetic field
\cite{DONG1,DONG2,RABHI,REZAEI,CHAKRABARTY,BRODERICK,MALLICK}. The
versatility of this formulation allows the inclusion of the
intrinsic magnetic moments in a covariant way. Due to the strength
of the baryon-meson couplings, the mean field approximation (MFA) is
usually employed. Within this approach the meson fields are replaced
by their expectation values and assimilated into a quasiparticle
picture of the baryons. Finally the meson mean values are obtained
by solving the classical meson equations taking as sources the
baryonic currents. This scheme is conceptually clear and easy to
implement.

The  propagators of charged particles in external magnetic fields
have been analyzed from different points of view
\cite{SCHWINGER,RITUS,PATKOS,KUZNETSOV,A&D2016}. A first attempt
to include the full effect of the intrinsic magnetic moments of a
Dirac particle propagating in a dense hadronic environment has
been presented in\cite{A&D2016}. However, it uses a mixed
representation where position and momentum variables are bound
together. The coherence of the approach was tested by evaluating
typical currents and densities in nuclear matter.\\
In the present work we try to improve that formalism, by
presenting a representation in terms of only momentum coordinates.
Obviously this fact makes easier the application of diagrammatic
procedures. In this sense, our results could be useful to complement
recent studies \cite{AYALA, CHEOUN, MANDAL}. \\
The propagator of Dirac particles subject to a uniform external
magnetic field, which includes the full effect of its magnetic
moments is used to evaluate corrections to the pion propagator in a
dense nuclear environment. In particular we evaluate the meson
polarization in conditions appropriate to defining the effective
pion mass. We analyze a wide range of densities $0 < n_B < 3 n_0$,
with $n_0$ being the saturation density of nuclear matter, and we
concentrate on very strong magnetic fields $B \geq 10^{17}$ G. Two
different isospin compositions of nuclear matter are considered,
pure neutron matter and symmetric nuclear matter.\\
The behavior of the pion polarization, and particularly of its
effective mass have recently been studied for low matter density and
a wide range of magnetic intensities \cite{COLUCCI,ADHYA}. However,
the intrinsic nucleon magnetic moments are not considered in these
works. In contrast, in our analysis the variation of the neutral
pion mass in pure neutron matter is an effect exclusively due to the
neutron magnetic moment.

This work is organized as follows. In the next section a summary of
the findings of Ref. \cite{A&D2016} is presented and further
development is made to derive a four-momentum representation for the
propagator. Since we are interested in evaluating the effects on the
pion propagation, we give a brief overview of the Green function for
a charged spin-zero meson in Sec. III. The evaluation of the
in-medium pion polarization and the definition of its effective mass
are given in Sec. IV. We devote Sec. V to the discussion of the
results. Finally, the conclusions are shown in Sec. VI.  Certain
details of the mathematical elaboration are transferred to the
Appendixes.

\section{In-medium propagator of a Dirac field with intrinsic magnetic
moment}\label{Sec1}

A preliminary version of the results of this section was presented
in Ref. \cite{A&D2016}. For the sake of completeness, we give here
an overview of the procedure.\\
The interaction of a spin $1/2$-fermion with a uniform magnetic
field is described by the Lagrangian density
\begin{eqnarray}
\mathcal{L}&=&\bar{\Psi}\Big[\gamma_\mu\left(i\,\partial^\mu-q\,
A^\mu\right)-m+\frac{\kappa}{2}\,\sigma^{\mu\nu}\,\mathcal{F}_{\mu\nu}\Big]\Psi\label{Lagran0}
\end{eqnarray}
where $\mathcal{F}^{\mu\nu}=\partial^\mu\,A^\nu-\partial^\nu\,A^\mu$
 and $\sigma^{\mu\nu}=i\,[\gamma^\mu, \gamma^\nu]/2$. For simplicity the
case of a uniform external magnetic field of magnitude $B$ along the
z-axis is considered, for which  $A^\mu=g^\mu_2  B x$.

For charged particles of positive energies $E_{n s}$, an exact
solution in mixed position and momentum coordinates can be written
as
\[    \phi_{n s p_y p_z}^{(+)}(\xi,y,z)= e^{i(p_y y+p_z
z)}\,e^{-\xi^2/2}\,u_{n s p_z}(\xi)\] where the index $s=1,-1$
stands for the spin projection along the magnetic field direction
and $n \geq 1$ denotes the quantized Landau levels.  We have also
used
\begin{eqnarray}
u_{n s p_z}(\xi)= N_{n s} \left(
\begin{array}{c}
H_n(\xi)\\ \\\frac{2\,n\,s\,p_z\,\sqrt{q B}\,i}{(\Delta_n+s\,m)\,(E_{n s}+s\,\Delta_n-\kappa B)}\,H_{n-1}(\xi)\\ \\
\frac{p_z}{E_{n s}+s\,\Delta_n-\kappa B}\, H_n(\xi)\\
\\-\frac{2\,n\,s\,\sqrt{q B}\,i}
{\Delta_n+s\,m}\,H_{n-1}(\xi)\\
\end{array}
\right) \label{Spinorp}
\end{eqnarray}
and,
\begin{eqnarray} \xi&=&(-p_y + q B x)/\sqrt{q B} \\
\nonumber \\
\Delta_n&=&\sqrt{m^2+2 n q B}\\
\nonumber \\
E_{n s}&=&\sqrt{p_z^2+(\Delta_n-s\,\kappa B)^2}\label{Pspectra}\\
\nonumber \\
N_{n s}^2&=&\frac{\sqrt{q B/\pi}}{(2 \pi)^2\,2^{n+2}\,n!}
\frac{(\Delta_n+s\,m)\, (E_{n s}+s\,\Delta_n-\kappa
B)}{m\,(\Delta_n-s\,\kappa B)}
\end{eqnarray}
$H_n$ stands for the Hermite polynomials. \\
The minimum energy eigenstate corresponds to $\phi_{0 p_y
p_z}^{(+)}(\xi,y,z)= e^{i(p_y y+p_z z)}\,e^{-\xi^2/2}\,u_{0 p_z}$,
with
\begin{eqnarray}
u_{0 p_z}= N_0 \left(
\begin{array}{c}
1\\ \\0\\ \\ \frac{p_z}{E_0+m-\kappa B}\\ \\0\\
\end{array}
\right) \label{Spinorp0}\end{eqnarray} and
\begin{eqnarray}
E_0&=&\sqrt{p_z^2+(m-\kappa B)^2}\label{Pspectr0}\\
\nonumber \\
N_0^2&=&\frac{\sqrt{q B/\pi}}{2\,(2 \pi)^2} \frac{(E_0+m-\kappa
B)}{(m-\kappa B)}
\end{eqnarray}

The antiparticle states correspond to the eigenvalues $-E_{n s}$ and
have eigenfunctions
\[  \phi_{n
s p_y p_z}^{(-)}(\xi,y,z)= e^{-i(p_y y+p_z z)}\,e^{-\eta^2/2}\,v_{n
s p_z}(\eta)\] with
\begin{eqnarray}
v_{n s p_z}(\eta)=N_{n s} \left(
\begin{array}{c}
\frac{p_z}{E_{n s}+s\,\Delta_n-\kappa B}\,H_n(\eta)\\
\\ \frac{2\,n\,s\,\sqrt{q B}\,i}
{\Delta_n+s\,m}\,H_{n-1}(\eta)\\ \\
H_n(\eta)\\ \\\frac{-2\,n\,s\,p_z\,\sqrt{q B}\,i}{(\Delta_n+s\,m)\,(E_{n s} +s\,\Delta_n-\kappa B)}\,H_{n-1}(\eta)\\
\end{array}
\right)
\end{eqnarray}
where $\eta=(p_y + q B x)/\sqrt{q B}$ and  $n\geq1$. The special
case $n=0$ has energy $-E_0$ and wave function $\phi_{0 p_y
p_z}^{(-)}(\eta,y,z)= e^{-i(p_y y+p_z z)}\,e^{-\eta^2/2}\,v_{0 p_z}$
with
\begin{eqnarray}
v_{0 p_z}=N_0 \left(
\begin{array}{c}
\frac{p_z}{E_0+m-\kappa B}\\ \\0\\ \\
1\\ \\0\\
\end{array}
\right)
\end{eqnarray}

For neutral particles $(q=0)$, the results are simpler. The particle
states are described by
\[ \phi_{\vec{p} s}^{(+)}(\vec{r})= e^{i
\vec{p}.\vec{r}}\,u_{\vec{p} s}\] with
\begin{eqnarray}
u_{\vec{p} s}= N_{\vec{p} s} \left(
\begin{array}{c}
1 \\ \\\frac{-s\,(p_x + i p_y)\,p_z}{(\Delta+s\,m)\,(E_{\vec{p} s}+s\,\Delta-\kappa B)} \\ \\
\frac{p_z}{E_{\vec{p} s}+s\,\Delta-\kappa B} \\ \\\frac{s\,(p_x + i p_y)}{\Delta+s\,m} \\
\end{array}
\right) \label{Spinorn}
\end{eqnarray}
and
\begin{eqnarray}
E_{\vec{p} s}&=&\sqrt{p_z^2+(\Delta-s\,\kappa B)^2}\label{Nspectra}\\ \nonumber \\
\Delta&=&\sqrt{m^2+p^2_x+p^2_y}\\ \nonumber \\N_{\vec{p}
s}^2&=&\frac{1}{4\,(2 \pi)^3} \frac{(\Delta+s\,m)\,(E_{\vec{p}
s}+s\,\Delta-\kappa B)} {m\,(\Delta-s\,\kappa B)}.
\end{eqnarray}
On the other hand, the antiparticle states have energies
$-E_{\vec{p} s}$, and eigenfunctions $\phi_{\vec{p} s}^{(-)
}(\vec{r})= e^{-i\vec{p}.\vec{r}}\,v_{\vec{p} s}$ with
\begin{eqnarray}
v_{\vec{p} s}=N_{\vec{p} s} \left(
\begin{array}{c}
\frac{p_z}{E_{\vec{p} s}+s\,\Delta-\kappa B} \\ \\
\frac{s\,(p_x + i p_y)}{\Delta+s\,m} \\ \\
1 \\ \\ \frac{- s\,(p_x + i p_y)\,p_z}{(\Delta+s\,m)\,(E_{\vec{p} s}+s\,\Delta-\kappa B)} \\
\end{array}
\right)
\end{eqnarray}

In the next step, we make a canonical expansion of the fermion
quantum fields   using the eigenfunctions just described. These
fields are used to evaluate the in-medium causal propagator
\cite{RITUS}
\[
i\, G_{\alpha \beta}(x',x)= <T \Psi_\alpha(x')
\bar{\Psi}_\beta(x)> \label{Green}
\]
Here the angular brackets must be regarded as a statistical mean
value, as obtained for instance, by evaluating the trace with the
density matrix of the system.\\
Using such a procedure, a mixed coordinates representation has been
obtained  for the covariant propagator of a charged Dirac particle
\cite{A&D2016}
\begin{eqnarray}
G_{\alpha \beta}(t',\vec{r}\,',t,\vec{r})&&=  \sqrt{\frac{q
B}{\pi}} \int \frac{dp_0\,dp_y\,dp_z}{(2 \pi)^3} e^{-i
p_0\,(t'-t)}\, e^{i[p_y (y'-y)+p_z (z'-z)]}\,e^{-(\xi'^2+\xi^2)/2}
\nonumber \\
&&\bigg \{\Lambda^0_{\alpha
\beta}\left[\frac{1}{p_0^2-E_0^2+i\epsilon}+ 2
\pi\,i\,n_F(p_0)\,\delta(p_0^2-E_0^2)\right]+ \sum_{n,s}
\frac{\Delta_n+s\,m}{2^{n+1}\,n!\,\Delta_n} \Lambda^{n s }_{\alpha
\beta}(\xi',\xi)
\nonumber \\
&&\times \left[\frac{1}{p_0^2-E_{n s}^2+i\epsilon}+2
\pi\,i\,n_F(p_0)\,\delta(p_0^2-E_{n s}^2)\right]\bigg \}
\label{GP}\end{eqnarray}
 where
\begin{eqnarray}
\Lambda^0&=&\left( \not \! u+m-\kappa\,B\right) \Pi^{(+)} \nonumber \\
\Lambda^{n s}&=&\Big[(\not \! u+s \Delta_n-\kappa B)\, H_n(\xi')+i
\, \frac{m - s \Delta_n}{\sqrt{q B}} \,
(\not \! u-s \Delta_n+\kappa B)\, \gamma^1 \,H_{n-1}(\xi')\Big ] \nonumber \\
&&\times \left[\Pi^{(+)} H_n(\xi)+i \, \frac{m - s
\Delta_n}{\sqrt{q B}}\, \gamma^1\, \Pi^{(-)} H_{n-1}(\xi)\right]
\nonumber \end{eqnarray}
and $\not \! u=p_0\gamma^0-p_z\gamma^3$,  $\Pi^{(\pm)}=(1\pm i\gamma^1\gamma^2)/2$, $\xi'=(-p_y + q B x')/\sqrt{q B}$. \\
For the neutral fermions it is
\begin{eqnarray}
G_{\alpha \beta}(x',x)= \sum_s \int \frac{dp^4}{(2 \pi)^4} e^{-i
p^\mu\,(x_\mu '-x_\mu)} \Lambda^s_{\alpha
\beta}\left[\frac{1}{p_0^2-E_{\vec{p} s}^2+i\epsilon}+ 2
\pi\,i\,n_F(p_0)\,\delta(p_0^2-E_{\vec{p} s}^2)\right]
\label{GN1}\end{eqnarray}
 where
\begin{eqnarray}
\Lambda^s=\frac{ s}{2 \Delta}i\; \gamma^1 \gamma^2\left[ \not \!
u+ i \gamma^1 \gamma^2 (s \Delta-\kappa B)\right] \left( \not \!
v+m+ i s \Delta \gamma^1 \gamma^2\right) \label{GN2}
\end{eqnarray}
and the notation $\not \! v=-p_x \, \gamma^1-p_y\, \gamma^2$ is
introduced.

 The formal difference between the results
for the neutral and charged fermionsmust be noted. In the first case
a pure momentum representation can be easily extracted from
Eq.(\ref{GN1}). This is not the case for Eq. (\ref{GP}), from which
a mixed-coordinate $(p_0,p_y,p_z;x,x')$ representation can be
deduced at most. This fact is related to the particular gauge chosen
for the electromagnetic field. Obviously, this is an undesirable
flaw for some specific
applications, for instance, diagrammatic expansions.\\
The problem of the gauge invariance of fermion propagators has been
discussed long time ago \cite{SCHWINGER}. By following such studies,
a transformation is applied to Eq. (\ref{GP}) which leads to the
following decomposition
\begin{equation}
G(x',x)=e^{i \Phi } \int \frac{d^4 p}{(2 \pi)^4} e^{-i
p^\mu\,(x'_\mu-x_\mu)} \left[ G_0(p)+\sum_{n,s} G_{n,s} (p)\right]
\label{DefGP0}
\end{equation}
with
\begin{eqnarray}
G_0(p)&=&2 e^{-p_\bot^2/q B}\Lambda^0
\left[\frac{1}{p_0^2-E_0^2+i\epsilon}+ 2
\pi\,i\,n_F(p_0)\,\delta(p_0^2-E_0^2)\right] \label{DefGP1}
\end{eqnarray}
\begin{eqnarray}
G_{n s}(p)&=&(-1)^n e^{-p_\bot^2/q B}\frac{\Delta_n+s m}{
\Delta_n}\Big\{( \not \! u-\kappa B+s \Delta_n) \Pi^{(+)} L_n(2
p_\bot^2/q B)-
\nonumber\\
&&( \not \! u+\kappa B-s \Delta_n) \Pi^{(-)} \frac{s \Delta_n-m}{s
\Delta_n+m} L_{n-1}(2 p_\bot^2/q B)+
\nonumber \\
&&\left[ \not \! u+i \gamma_1 \gamma_2 (s \Delta_n-\kappa B)\right]
i \gamma^1 \gamma^2 \not \! v \frac{s \Delta_n-
m}{2\,p_\bot^2}\left[ L_n(2 p_\bot^2/q B)-L_{n-1}(2 p_\bot^2/q
B)\right]\Big\}
\nonumber \\
&&\times \left[\frac{1}{p_0^2-E_{n s}^2+i\epsilon}+2
\pi\,i\,n_F(p_0)\,\delta(p_0^2-E_{n s}^2)\right] \label{DefGP2}
\end{eqnarray}
here $L_m$ stands for the Laguerre polynomial of order $m$, and
$p_\bot^2=p_x^2+p_y^2$ is used. The phase factor $\Phi=q
B(x+x')(y'-y)/2$ embodies the gauge fixing. For mathematical details
see  Appendix A.

\section{In-medium propagator of a charged scalar Bose field}

The covariant propagator of a charged scalar field in the presence
of an external magnetic field has been studied in the past,
including the method of eigenfunctions expansion \cite{PATKOS}. We
present here a procedure which renders the propagator into a
four-dimensional momentum representation.
\\
The meson field $\phi(x)$ interacting with a electromagnetic field
$A_\mu(x)$ is described by the lagrangian
\[\mathcal{L}=\left(\partial^\mu-i e A^\mu \right)\varphi^\dag \left(\partial_\mu+i e A_\mu
\right)\varphi-m^2 \varphi^\dag \varphi
\]
We choose the gauge as in Section \ref{Sec1}. The eigenfunctions
 are
\[
\varphi_n(x)=\mathcal{N} e^{-i(\omega_n t-p_y y-p_z z)}
e^{-\xi^2/2} H_n(\xi)
\]
with $\omega_n=\sqrt{m^2+p_z^2+2(n+1)q B}$, $\xi=\sqrt{q
B}(x-p_y/q B)$, and $ \mathcal{N}^2=\sqrt{q B/\pi}/2^n n!$ An
expansion of the quantum field is proposed as
\[\phi(x)=\sum_{n,l}\int \frac{dp_z}{2 \pi \omega_n}\left[\varphi_{n l} a_{n l}(p_z)+ \varphi_{n l}^\ast b_{n l}^\dag(p_z)\right]
\]
with canonical commutation relations for the creation and
destruction operators. Using the standard definition of the
propagator $i\, \Delta(x,x') = <T \phi (x) \phi^\dag (x')>$, where
angular brackets stand for a statistical expectation value, we
obtain
\begin{eqnarray}
\Delta(x,x')&=&\sum_n \mathcal{N}^2 \int \frac{dp_0 dp_y dp_z}{(2
\pi)^3} e^{-i p_0(t-t')+i p_y(y-y')+i p_z(z-z')} H_n^(\xi)
H_n^(\xi')e^{-(\xi^2+\xi'^2)/2}
\nonumber \\
&&\times \left[\frac{1}{p_0^2-\omega_n^2+i \varepsilon}+2 \pi i
\delta(p_0^2-\omega_n^2) n_B(p_0) \right] \nonumber
\end{eqnarray}
 where
$n_B$ is the Bose distribution function, and we use the identity
\[ \frac{1}{2 \omega_n}\left[\Theta(t'-t) e^{i \omega_n(t-t')}+\Theta(t-t') e^{-i \omega_n(t-t')}
\right]=\frac{i}{2 \pi}\int \frac{dp_0}{2 \pi} \frac{e^{-i
p_0(t-t')}}{p_0^2-\omega_n^2+i \varepsilon}
\]
in order to unify particle and antiparticle notation.\\
In the first place we perform the integration over $p_y$,
\[ \int dp_y \mathcal{N}^2 e^{i p_y(y-y')} e^{-(\xi^2+\xi'^2)/2} H_n(\xi)
H_n(\xi')= q B e^{i \Phi} e^{- q B R^2/4} L_n(q B R^2/2)
\]
with the help of Eq. 7.377 of Ref. \cite{G&R}, where
$R=\sqrt{(x-x')^2+(y-y')^2}$ is used. The right-hand side can be
rewritten in terms of a bidimensional momentum integral by using Eq.
(\ref{App6}). Hence, we finally obtain
\begin{eqnarray}
\Delta(x,x')&=&e^{i \Phi} \int\frac{d^4p}{(2 \pi)^4} e^{-i p^\mu
(x-x')_\mu}\Delta(p)\nonumber
\end{eqnarray}
\begin{eqnarray}
\Delta(p)&=&2 \sum_n (-1)^n e^{-p_\bot^2/q B} L_n(2 p_\bot^2/q B)
\left[\frac{1}{p_0^2-\omega_n^2+i \varepsilon}+2 \pi i
\delta(p_0^2-\omega_n^2) n_B(p_0) \right] \nonumber
\end{eqnarray}
The quantities $\Phi$ and $p_\bot$ were defined at the end of Sec.
\ref{Sec1}.

\section{Pion effective mass in the nuclear medium under a uniform magnetic field}

In this section we consider the hadronic interaction in the
presence of a uniform external magnetic field. It is described by
a QHD model, where baryons interact with pions and neutral mesons
$\sigma$ and $\omega$. The langragian density is
\begin{eqnarray}
\mathcal{L}&=&\sum_{a=n,p}
\bar{\Psi}^a\Big[\gamma_\mu\left(i\,\partial^\mu-q_a\, A^\mu+g_w
\omega^\mu-\frac{g_A}{2 f_\pi} \gamma_5 \bm{\tau} \cdot
\partial^\mu \bm{\phi}-\frac{1}{4 f_\pi^2}\bm{\tau} \cdot \bm{\phi} \times \partial^\mu \bm{\phi}\right)
\nonumber \\ &&
-m_0+g_s\sigma+\frac{\kappa}{2}\,\sigma^{\mu\nu}\,\mathcal{F}_{\mu\nu}\Big]\Psi^a
\nonumber \\
&-&\frac{1}{4}\,\mathcal{F}_{\mu\nu}\,\mathcal{F}^{\mu\nu}
+\frac{1}{2}\,(\partial_\mu\sigma\,\partial^\mu\sigma-m_\sigma^2\,\sigma^2)
-\frac{1}{4}\,\Omega_{\mu\nu}\,\Omega^{\mu\nu}+
\frac{1}{2}\,m_\omega^2\,\omega_\mu\,\omega^\mu+\frac{1}{2}\,\partial_\mu\phi^0\,\partial^\mu\phi^0\
\nonumber
\\&& +\left(\partial^\mu-i e A^\mu \right)\phi^-
\left(\partial_\mu+i e A_\mu \right)\phi^+-\frac{1}{2} m_\pi^2
\bm{\phi}\cdot \bm{\phi}\label{Lagran}
\end{eqnarray}

where
$\Omega^{\mu\nu}=\partial^\mu\,\omega^\nu-\partial^\nu\,\omega^\mu$
and only protons and neutrons have been included. In this approach,
the fundamental state of matter is given by a MFA, which is
equivalent to including the tadpole diagram (see Fig. 1a) in a
self-consistent solution but neglecting divergent contributions
coming from the Dirac sea. At this step it is assumed that meson
propagation is not modified by the hadronic interaction. The effect
of the magnetic field, instead, is fully included for both meson and
nucleon propagators. This means that fermionic lines in the diagram
of Fig. 1(a) correspond to either Eqs. (\ref{GN1})-(\ref{GN2}) or
Eqs.
(\ref{DefGP1})-(\ref{DefGP2}).\\
It can be verified that pions do not contribute to the tadpole
diagram, since the pion-nucleon vertices depend on the transferred
pion momentum. Furthermore the neutral mesons $\sigma, \, \omega$
and $\pi^0$ are not affected directly by the magnetic field.\\
It is well known that in QHD models the  MFA leads to a
quasiparticle picture for nucleons, where the mass and energy
spectra are modified, according to $m=m_0-g_s \sigma_0$, $p_0=g_w
w_0\pm E$, with $E$ being one of the eigenvalues shown in Eqs.
(\ref{Pspectra}), (\ref{Pspectr0}), or (\ref{Nspectra}). The
quantities $\sigma_0, \, w_0$ correspond to the in-medium
expectation values of the $\sigma$ and timelike component of
$\omega$ mesons \cite{SW} $w_0=g_\omega  n_B/m_\omega^2$ and
$\sigma_0=g_\sigma n_s/m_\sigma^2$. The baryonic number ($n_B$) and
scalar ($n_s$) densities can be decomposed into their neutron and
proton components
\begin{eqnarray}
n_B^{(n)}&=&\sum_s \int \frac{dp^3}{(2 \pi)^3}
 \left[ n_F(E_{\vec{p} s})-n_F(-E_{\vec{p} s})\right]\nonumber
\\
n_B^{(p)}&=&\frac{q B}{2 \pi^2} \int dp_z
\Big\{\left[n_F(E_0)-n_F(-E_0)\right]+\sum_{n,s} \left[n_F(E_{n
s})-n_F(-E_{n s})\right]\Big\}
\nonumber \\
n_s^{(n)}&=&\sum_s\int \frac{dp^3}{(2 \pi)^3}
 \frac{\Delta+s\,\kappa_n B}{E_{\vec{p}
s}\,\Delta}\left[n_F(E_{\vec{p} s})+n_F(-E_{\vec{p}
s})\right]
\nonumber \\
n_s^{(n)}&=&\frac{q B}{2 \pi^2} \int dp_z
\Big\{\frac{m+\kappa_p\,B}{E_0}\left[ n_F(E_0)+n_F(-E_0)\right]
\nonumber \\
&& +m\,\sum_{n, s} \frac{\Delta_n+s\,\kappa_p B}{E_{n
s}\,\Delta_n}[n_F(E_{n s})+n_F(-E_{n s})] \Big\} \nonumber
\end{eqnarray}

Therefore, at the end of the calculations, we formally recover
similar expressions for the nucleon propagators as given in Sec.
\ref{Sec1}, but with the following modifications: {\it i}) the
nucleon vacuum mass is replaced by the in-medium effective mass, and
{\it ii}) the variable $p_0$ must be replaced by
$\tilde{p}_0=p_0-g_w w$. For practical applications, the last point
is equivalent to replacing the thermodynamical chemical potential
$\mu$ with the effective potential $\tilde{\mu}=\mu-g_w w$.

In the next step we study the effects of the hadronic interaction on
the meson properties. In particular we consider the modification of
the pion mass. For this purpose we evaluate the pion polarization
insertion due to Figs. 1(b) and 1(c) of Fig.1 corresponding to
the one-loop approximation.\\
The diagram in Fig. 1(b) comes from the Weinberg-Tomozawa term and
corresponds to first-order correction. It gives nonzero
contributions only for the nondiagonal channel (1,2) of the
hermitian pion fields
\begin{equation}
\Pi_{a b}^{WT}(p)=-\frac{\varepsilon_{3 a b}}{2 f_\pi^2}\, p_\mu
\sum_c \tau_3^{c c} \int \frac{d^4q}{(2 \pi)^4} Tr\left\{\gamma^\mu
G^{(c)}(q) \right\}\label{PolWT0}
\end{equation}
where the sum runs over protons ($c=1$) and neutrons ($c=2$).
Using the nucleon propagators constructed in the MFA and
specializing for the conditions of interest for our calculations,
we finally obtain
\begin{equation}
\Pi_{\pm}^{WT}(p_0,{\bm p}=0)=\pm  \frac{p_0}{2 f_\pi^2}\,
\left(n_B^{(n)}-n_B^{(p)}\right) \label{PolWT}
\end{equation}
whereas $\Pi_0^{WT}(p)=0$.

The diagram in Fig. 1(c) corresponds to the pseudovector one-pion
vertex (OPV), it is a second order correction
\begin{equation}
i \Pi^{OPV}(p)=\left( \frac{g_A}{2 f_\pi}\right)^2 p_\mu p_\nu \int
\frac{d^4q}{(2 \pi)^4} Tr\left\{\gamma^\mu \gamma_5 G^{(a)}(q)
\gamma^\nu \gamma_5 G^{(b)}(q-p)\right\} \label{PolOPE}
\end{equation}
It is understood that for the neutral pion a sum over $a=b$ must be
done, for the positively charged pion is $a=p, \, b=n$, and finally
$a=n, \, b=p$ corresponds to the negatively charged pion. Explicit
expressions for the polarizations $\Pi^{OPV}_0, \Pi^{OPV}_\pm$
evaluated at ${\bf p}=0$ are shown in Appendix B.

In a Dyson-Schwinger approach, the poles of the pion propagator are
modified by the polarization insertion. For charged pions and for
each Landau level they are given by $p_0^2-\omega_n^2-\Pi_\pm(p)=0$.
For neutral pions instead, they are defined by $p_0^2-(m_\pi^2+{\bm
p}^2)-\Pi_0(p)=0$. Hence, the in-medium effective mass is defined as
the solutions of the following equation for $p_0$
\begin{equation}
p_0^2-m_\pi^2-\Pi(p_0,{\bm p}=0)=0 \label{MassDef}
\end{equation}
It must be noted that  the term $2 n q B$ coming from the quantized
Landau states for the charged pions has not been included in this
definition.\\ As it was already mentioned, for charged pions the
polarization insertion in Eq. (\ref{MassDef}) is a sum
$\Pi_{\pm}^{WT}+\Pi^{OPV}_\pm$ whereas for the neutral pion there is
only one contribution $\Pi^{OPV}_0$.

\section{Results and discussion}

In this section we solve Eq. ({\ref{MassDef}) for different
situations of physical interest. We consider here very strong
magnetic fields $10^{16}-10^{19}$ G, and matter at zero temperature
and baryonic densities below $0.45$ fm$^{-3}$. We also take the
isospin composition of matter as a variable and examine two
different situations: {\it i}) symmetric nuclear matter
$n_B^{(p)}=n_B^{(n)}$, and {\it ii}) pure neutron matter
$n_B^{(p)}=0$. In the first case there is no contribution from
$\Pi_{\pm}^{WT}$, as can be seen from Eq. (\ref{PolWT}).\\

In first place we take a look of the thermodynamical state of matter
at zero temperature,  within the MFA.  The results obtained are used
to evaluate the medium-dependent parameters of the fermionic
propagators. \\ As a second step, the polarization insertion for
pions is constructed and its effective mass is examined.

\subsection{Thermodynamics of the magnetized hadronic medium}

For a given magnetic intensity, baryonic density and isospin
composition, the equilibrium state of matter corresponds to a
minimum of the energy of the system. In this state, each isospin
component acquires a global spin polarization, induced by the
external magnetic field. Furthermore, the system exhibits a weak
magnetization. \\
In this section we examine the thermodynamical properties of the
equilibrium state.

  In Fig. 2 we
present the energy per particle (with subtraction of the nucleon
mass $m_0$) as a function of the density for several magnetic
intensities. In symmetric matter a minimum or saturation point is
found, whose energy decreases with the intensity of the magnetic
field. The fact that nuclear matter is more strongly bound as the
magnitude of the external field grows has been remarked on in
different studies \cite{DONG1,AB&V,RABHI,REZAEI}. For pure neutron
matter, instead, a monotonous increase is found. It deserves to be
mentioned that at low densities neutron matter becomes bound, a
feature emphasized as $B$ increases \cite{DONG2,AB&V}.

Next, in Fig. 3, we analyze the spin polarization of each isospin
component as a function of the density. The quantity
$W^{(a)}=(n^{(a)}_{\text{up}}-n^{(a)}_{\text{down}})/n_B^{(a)}$ with
$a=p, n$, gives a statistical measure of the fraction of particles
with spin polarized in the same direction of the field or in
opposition to it. For $B >10^{16} G $ both isospin components are
completely polarized at very low densities. There is an abrupt
change of polarization for the weaker intensities, while the plateau
of complete polarization is extended in density as the external
field grows. In this sense, the response of the proton component
seems to be more intense than the neutron one. As a special
situation, it can be seen that the curve corresponding to $W^{(p)}$
for symmetric matter at $B=10^{18}$ G shows several irregularities
due to the thresholds in the occupation of different Landau
levels.\\
The case $B=10^{16}$ G has not been included in Figs.2 and 3, since
it is indistinguishable from the $B=10^{17}$ G curve for the scale
shown. In both figures there is an apparent difference in the
qualitative behavior of the $B=5\times 10^{18}$ G curves, which can
be attributed to the intrinsic magnetic moments
\cite{RABHI,DONG1,REZAEI}, insofar as $\kappa_a B \sim 1$.

The effective chemical potential for neutrons and protons as a
function of density is exhibited in Fig. 4. It can be appreciated
that $\tilde{\mu}$ is a decreasing function, even though the
thermodynamical potential $\mu$ increases monotonously, as it
corresponds to a thermodynamical equilibrium state. This is a
consequence of the faster growth of the mean field value $\omega_0$.
Only the case $B=5 \times 10^{18}$ G is shown in this figure, but a
similar trend is obtained for other intensities.

The equation of state of hadronic matter in the presence of an
external uniform magnetic field has been widely discussed. In
particular, Ref. \cite{BRODERICK} makes an exhaustive analysis of
the effects of the anomalous magnetic moments (AMM) in matter
composed of nucleons, light mesons and electrons. One of the results
shown there, is the relevance of the AMM for sufficiently high
magnetic intensities. These results apparently contradict the
conclusions of Ref. \cite{Chanta1}. However, a direct comparison is
not fair for several reasons: {\it i})  A system of structureless
charged fermions interacting solely with the electromagnetic field
is considered in Ref. \cite{Chanta1}. In such an approach the AMM is
obtained from an expansion of the fermion self-energy. In hadronic
physics instead, the AMM are taken as constants determined mainly by
the quark structure of hadrons. Hence, even neutral fermions exhibit
nonzero AMM. {\it ii}) The equation of state is evaluated in Ref.
\cite{Chanta1} for only one species of charged fermions. This
description can hardly be applied to realistic situations of
hadronic physics. By way of illustration neutron star matter can be
considered. In such a case, the Coulomb repulsion among charged
baryons is modified by the interaction with neutral ones.
Furthermore, charged leptons are necessary to locally fulfill the
requirement of charge neutrality. {\it iii}) The explicit
calculations of the equation of state shown in \cite{Chanta1} are
particularly misleading for hadronic physics, as the set of
numerical values used there causes a loss of generality. For
instance, the fermion mass fixed at m=0.5 MeV is almost irrelevant
for baryons, and consequently the chemical potential $\mu$=10
MeV$\simeq$0.05 fm$^{-1}$ corresponds, at zero temperature, to
unphysical high densities (see Fig. 4).

\subsection{In medium pionic mass}

The effective masses of the baryons and their chemical potentials
obtained in the MFA are taken as input for the propagators
(\ref{GN1})-(\ref{GN2}) and (\ref{DefGP0})-(\ref{DefGP2}). In turn,
they are used to evaluate the polarization insertions (\ref{PolWT0})
and (\ref{PolOPE}) as functions of the density and the magnetic
intensity.

In Figs. 5 and 6 the solutions of Eq. (\ref{MassDef}) are examined
in terms of the particle density for several field intensities. The
case $B=10^{19}$ G is also included, because all the effects
discussed are enlarged for this extremely strong field. It must be
emphasized that we do not include vacuum corrections and hadrons are
regarded as elementary degrees of freedom.

 The density dependence of the pion mass in neutral matter
is exhibited in Fig.5. The charged pions (Fig. 5a) receive
contributions  from $\Pi_{\pm}^{WT}$, and from the OPV between
neutrons and protons in the Fermi sea. The first term does not
depend explicitly on the magnetic intensity, but only through the
particle density. Therefore the same curve corresponds to different
values of $B$ (dark lines). In fact, it can be shown that neglecting
corrections from the OPV, the effective pion masses can be written
as $m_{\pm}^*/m_\pi=\sqrt{1+(a r)^2}\pm a r$, where $a=n_0/4 f_\pi^2
m_\pi$ and the relative density $r=n_B/n_0$ is introduced. When the
full term is considered, a weak dependence on $B$ emerges. Only for
the strongest intensity $B=10^{19}$ G (gray lines)do the differences
become appreciable.
\\
The neutral pion receives a contribution only from the neutrons
through $\Pi_0^{OPV}$. The composition of this term is shown in
detail in Eqs. (\ref{AppB1})-(\ref{AppB3}). The factor
$\Theta(\tilde{\mu}_n-|{\cal M}_s|)$ disfavors the spin-up
contribution by reducing the integration domain of Eqs.
(\ref{AppB2}) and (\ref{AppB3}), which numerically are both
positive. As the spin polarization drops abruptly at low densities
for $B=10^{17}-10^{18}$ G (see Fig. 3b), $\Pi^{OPV}_0$ partially
loses the spin-up contribution at lower densities as compared to
higher values of $B$. This explains the weaker growth at low and
medium densities observed in Fig. 5b for $B<5 \times 10^{18}$ G.
This effect is accentuated as the density increases because
$\tilde{\mu}_n$ shows a decreasing behavior (see Fig. 4). On the
other hand the coefficient $\kappa_n B$ in the second term of Eq.
(\ref{AppB1}), reduces drastically the contribution of Eq.
(\ref{AppB3}) when $B<5 \times 10^{18}$ G. From the numerical
analysis it is found that Eq. (\ref{AppB3}) increases with density
and for a given value of $n_B$ is significantly greater than Eq.
(\ref{AppB2}). As a consequence we observe a high slope for
$B=10^{19}$ G, a moderate slope for $B=5\times 10^{18}$ G, and a
tiny slope for the remaining cases.\\
It must be pointed out that the dependence on the magnetic field
shown in Fig. 5b is an exclusive consequence of the neutron magnetic
moment. If $\kappa_n=0$, all the curves will coincide.

Symmetric nuclear matter is considered in Fig. 6,  in such
conditions the polarization is solely due to the OPV. In
particular for the neutral pion (Fig. 6c) there is a sum of
independent proton and neutron terms. The neutron contribution
produces a smooth dependence as discussed above. In fact, the
results for the neutral pion mass in neutron or symmetric nuclear
matter are qualitatively similar. The rate of growth is slightly
more pronounced for the latter case, with the exception of $B=10^{19}$ G.\\
For the charged pions, there is a mix of neutron and proton terms in
each diagram. In contrast to the previous case, there are some
pronounced irregularities, related to the occupation of the discrete
Landau levels, which are emphasized as the magnetic field increases.
The case of the negatively charged pion is examined in Fig. 6a. For
the lowest magnetic intensities, the proton levels are gradually and
almost smoothly occupied as the density increases, reaching $n=120$
for $B= 10^{17}$ G and $n=12$ for $B= 10^{18}$ G. As a consequence
the effective mass increases monotonously in the first case and a
mild oscillatory behavior appears in the second case. A drastic
change is observed for $B=5 \times 10^{18}$ G. There are abrupt
modifications of the slope at the densities $n_B/n_0\simeq 0.6, 1.4,
1.6$, and $2.9$. To be more precise, the curve exhibits at these
points local maxima followed by a fast increase.  At these densities
preciselythe opening of the Landau levels $n=1,\, s=1$, $n=2, \,
s=1$, the threshold of the proton depolarization (see Fig. 3a), and
the beginning of the population of the $n=2,\, s=-1$ level,
respectively, take place . In contrast to the previous cases, these
changes take place at relatively greater densities (with higher
Fermi momentum), hence the effect is amplified. For the most intense
field considered here, $B= 10^{19}$ G, a drop of roughly $4 \, \%$
occurs at $n_B/n_0\simeq 1.6$, followed by a further increase. At
this point the Landau level $n=1$ becomes
available.\\
In fact the same description, but at a considerably minor scale,
holds for $B=5 \times 10^{18}$ G, $n_B/n_0\simeq 0.6$ where the
proton phase is completely polarized and the population of the
first excited Landau level is initiated.\\
In regard to the positively charged pion, it must be noted that some
pairs of terms in Eq.(\ref{RePiPlus}) contribute with opposite
signs, in contrast to the result for Eq.(\ref{RePiMinus}). As a
consequence, the same causes have qualitatively different
manifestations for the results  shown in Fig. 6b, and for those just
discussed for Fig. 6a. For instance, when $B= 10^{19}$ G, a jump in
the effective mass is present at $n_B/n_0\simeq 1.6$.  For densities
above or below this point, the effective mass is increasing at the
beginning, then it stabilizes and eventually decreases. This effect
is due to the cancellation of pairs of terms growing with density
but with opposite signs. A similar sketch is obtained for $B=5
\times 10^{18}$ G, but with significative points at $n_B/n_0\simeq
0.6,
1.4, 1.6, 2.9$.\\
For the two remaining values of $B$, a smooth behavior is
obtained, in agreement with the correspondent results shown in
Fig. 6a.\\
To complete this discussion, the effective pion mass obtained for
symmetric nuclear matter in the lowest Landau level approximation is
shown in Fig. 7. In this approach the summation over Landau levels
with $n \geq 1$ is neglected. From comparison with Fig. 6, we
conclude that, with exception of the very low density regime, this
approximation is not acceptable for intensities $B\leq 10^{18}$ G.
As $B$ is increased, the domain of completely polarized proton phase
is extended, and the approximation results are more adequate. For
instance, qualitative agreement is obtained until $n_B/n_0\sim 1.5$,
for $B=5 \times 10^{18}, \, 10^{19}$ G.

To end this section, we consider the imaginary part of the pion
polarization, which is related to the in-medium dynamical stability
of the particle. We have checked that the quantity $\tau= |\text{Im}
\, \Pi(p_0=m^*)|/m^*$ increases with the density and the magnetic
intensity. For all the ranges of densities and magnetic intensities
considered in our calculations, we have verified  that $\tau<0.12$.

\section{Conclusions}
In this work a covariant calculation of the propagator of  Dirac and
spin-zero Bose fields in the presence of a uniform external magnetic
field has been presented. The nonzero magnetic moment of the fermion
has been fully taken into account. The expressions found are valid
for finite temperature and density. Furthermore, the gauge dependent
contribution is reduced to a phase term as in the proper time
evaluation of \cite{SCHWINGER}. The propagators depend only on the
four-momentum, improving the results found in Ref. \cite{A&D2016}.
\\The formalism has been applied to define an effective mass of
the pion field, propagating in a dense nuclear medium at zero
temperature. The approach proposed neglects divergent contributions
from the Dirac sea, as it is a common practice in QHD calculations.
It must be taken into account that QHD models exhibit vacuum
instabilities \cite{Chanta2,Chanta3,Chanta4} which appear for
transferred momentum above $2-3$ GeV \cite{Chanta2}, when Dirac
contributions to the diagram in Fig. 1(c) are included. This result
questions the original interpretation of this kind of model as
realistic field theory description. A procedure for eliminating such
instabilities was presented in Ref. \cite{Chanta4}. The significant
fact that the model breaks down at length scales $0.2$ fm, smaller
than the nucleon size, could indicate the emergence of substructure
effects. Thus, QHD can be regarded as phenomenological descriptions
which include interactions and solution procedures in its
formulation. Therefore, it will be interesting to complement the
results shown with a correction which takes account of the structure
of hadrons.

The effective pion mass has been examined for two conditions of
interest in practical application: pure neutron matter and isospin
symmetric nuclear matter. Furthermore, we have focused on the domain
of very strong magnetic fields $B\geq 10^{17}$ G and have covered
particle densities below three times
the nuclear saturation density. \\
In most situations the effective mass increases with the density. As
the magnetic intensity grows, the behavior is marked by the
thresholds of the proton Landau levels, and the change of the spin
polarization of the nucleons. \\
Furthermore, by taking the imaginary part of the pion polarization
as a measure of its stability in the nuclear medium, we have found
that the neutral and positive pions are stable, and the negative
pion becomes slightly unstable for high densities and magnetic
intensities.

\section{Acknowledgements}
This work has been partially supported by the Consejo Nacional de
Investigaciones Cientificas y Tecnicas, Argentina.

\section{Appendix A}\label{AppA}

In order to derive Eqs. (\ref{DefGP0})-(\ref{DefGP2}) from Eq.
(\ref{GP}) we first integrate $p_y$ separately for the first and
second terms between curly brackets.\\
In the first case the relation
\begin{equation}\int \frac{dp_y}{2
\pi} e^{i p_y(y'-y)} e^{-(\xi^2+\xi'^2)/2}=\sqrt{\frac{q B}{4
\pi}} e^{i \Phi} e^{-q B [(x-x')^2+(y-y')^2]/4} \label{App1}
\end{equation}
is used, which follows from Eq. 17.23 (13) of Ref. \cite{G&R}.\\
For the next step the following relations will be useful
\begin{eqnarray}
\int_{-\pi}^{\pi} d\theta \, e^{i z \cos(\theta-\varphi)}=2 \pi
J_0(z) \label{App2} \\
\int_{-\pi}^{\pi} d\theta \,e^{\pm i \theta} e^{i z
\cos(\theta-\varphi)}=2 \pi i\,e^{\pm i \varphi} J_1(z)
\label{App3}
\end{eqnarray}
which can be deduced with the help of Eq. 8.511 (4) of Ref. \cite{G&R}.\\
The last exponential on the right hand side of Eq. (\ref{App1}) can
be expressed in terms of a bidimensional integral on the momentum
plane orthogonal to the external field
\[
e^{-q B [(x-x')^2+(y-y')^2]/4}=\frac{1}{\pi q B}\int_0^\infty
dp_\bot \, p_\bot e^{-p_\bot^2/q B} \int_{- \pi}^\pi d\theta \,
e^{i p_\bot R \cos (\theta-\varphi)}
\]
where $p_\bot=\sqrt{p_x^2+p_y^2}$, and
$R=\sqrt{(x-x')^2+(y-y')^2}$. Furthermore, $\theta, \, \varphi$
are the angular coordinates on the orthogonal plane of the vectors
$(p_x,p_y)$, and $(x'-x,y'-y)$ respectively. For this purpose Eq.
6.631(4) of Ref. \cite{G&R} and Eq. (\ref{App2}) have been
successively used. Thus, the relation
\[ \int \frac{dp_y}{2 \pi} e^{i p_y(y'-y)}
e^{-(\xi^2+\xi'^2)/2}= \sqrt{\frac{4 \pi}{q B}}\, \, e^{i \Phi}
\int \frac{dp_x dp_y}{(2 \pi)^2}e^{i[p_x(x'-x)+p_y(y'-y)]}
e^{-p_\bot^2/q B}
\]
is established.\\
On the other hand, by using Eq. 7.377 of Ref. \cite{G&R} it can be
shown that
\begin{eqnarray}
\int \frac{dp_y}{2 \pi} e^{i p_y(y'-y)} e^{-(\xi^2+\xi'^2)/2}
\Lambda^{n s}=\sqrt{\frac{q B}{4 \pi}}\,e^{i \Phi} e^{-q B R^2/4}
2^n n!
\nonumber\\
\Big\{ (\not \! u-\kappa B+s \Delta_n)\big[\Pi^{(+)} L_n +i
\,\gamma_1 \Pi^{(-)}\,\frac{m-s\Delta_n}{2 n}\, R \,e^{-i
\varphi}\, L_{n-1}^1 \big]+
\nonumber \\
(\not \! u+\kappa B-s \Delta_n)\frac{s\Delta_n - m}{s\Delta_n + m}
\big[\Pi^{(-)} L_{n-1}+ i \,\gamma_1\,\Pi^{(+)}\,
\frac{m+s\Delta_n}{2 n}\, R\,e^{i \varphi} \, L_{n-1}^1
\big]\Big\} \label{App4}
\end{eqnarray}
where the argument of all the Laguerre polynomials on the right-hand
side is $q B R^2/2$, and by definition $R \cos\varphi=x'-x,\;R
\sin\varphi=y'-y$. Use has been
made of the fact that Laguerre polynomials have definite parity.\\
Furthermore, the relations
\begin{eqnarray}
e^{-q B R^2/4} L_n\left(q B R^2/2\right)&=&\frac{(-1)^n}{\pi q B}
\int_0^\infty dp_\bot \, p_\bot e^{-p_\bot^2/q B} L_n\left(2
p_\bot^2/q B\right)
\nonumber \\
&&\times \int_{- \pi}^\pi d\theta \, e^{i p_\bot R \cos
(\theta-\varphi)} \label{App6}
\end{eqnarray}
\begin{eqnarray}
R e^{\pm i\varphi} e^{-q B R^2/4} L_n^1(q B
R^2/2)&=&\frac{(-1)^n}{2 \pi i}\left(\frac{2}{q B} \right)^2
\int_0^\infty dp_\bot \,p_\bot^2\,
 e^{-p_\bot^2/q B} L_n^1(2 p_\bot^2/q B)
 \nonumber \\
 &&\times \int_{- \pi}^\pi d\theta \,e^{\pm i \theta}\,e^{i p_\bot R \cos
(\theta-\varphi)}
\end{eqnarray}
are obtained from Eq. 7.421(4) of Ref. \cite{G&R} and Eqs.
(\ref{App2}), (\ref{App3}). \\
When the last two equations are inserted into Eq. (\ref{App4}) the
result
\begin{eqnarray}
\int \frac{dp_y}{2 \pi} e^{i p_y(y'-y)} e^{-(\xi^2+\xi'^2)/2}
\Lambda^{n s}=\sqrt{\frac{4 \pi}{q B}}\,e^{i \Phi} 2^n n! (-1)^n
\int \frac{dp_x dp_y}{(2 \pi)^2} e^{i[p_x(x'-x)+p_y(y'-y)]}
e^{-p_\bot^2/q B}
\nonumber \\
\Big\{ (\not \! u-\kappa B+s \Delta_n)\Pi^{(+)} L_n-(\not \!
u+\kappa B-s \Delta_n)\Pi^{(-)}\frac{s\Delta_n-m}{s\Delta_n+m}
L_{n-1}+
\nonumber \\
\frac{s \Delta_n-m}{n q B} p_\bot \Big[ (\not \! u-\kappa B+s
\Delta_n)\gamma^1 \Pi^{(-)} \, e^{-i \theta}-(\not \! u+\kappa B-s
\Delta_n)\gamma^1 \Pi^{(+)} \, e^{i \theta}\Big] L_{n-1}^1\Big\}
\end{eqnarray}
is obtained.\\
After some algebra on the last term between curly brackets and the
use of Eq. 8.971 (4) of Ref. \cite{G&R} to put $L_k^{(1)}$ in
terms of $L_k-L_{k-1}$, we obtain the final expression
\begin{eqnarray}
\int \frac{dp_y}{2 \pi} e^{i p_y(y'-y)} e^{-(\xi^2+\xi'^2)/2}
\Lambda^{n s}=\sqrt{\frac{4 \pi}{q B}}\,e^{i \Phi} 2^n n! (-1)^n
\int \frac{dp_x dp_y}{(2 \pi)^2} e^{i[p_x(x'-x)+p_y(y'-y)]}
e^{-p_\bot^2/q B}
\nonumber \\
\Big\{ (\not \! u-\kappa B+s \Delta_n)\Pi^{(+)} L_n-(\not \!
u+\kappa B-s \Delta_n)\Pi^{(-)}\frac{s\Delta_n-m}{s\Delta_n+m}
L_{n-1}+
\nonumber \\
\frac{s \Delta_n-m}{2 p_\bot^2}  \left[ \not \! u+(s
\Delta_n-\kappa B)i \gamma^1 \gamma^2\right] i\gamma^1 \gamma^2
\not \! v \, ( L_n-L_{n-1})  \Big\}
\end{eqnarray}

\section{Appendix B}\label{AppB}

Here we give explicit formulas for the one-pion exchange
contribution to the pion polarization at ${\bf p}=0$.\\
Since
\[ \Pi_0=\Pi_0^{(n)}+ \Pi_0^{(p)}\]
we have

\begin{eqnarray}
\text{Re} \, \Pi^{(n)}_0 (p_0)&=&\left(\frac{g_A}{4 \pi f_\pi}
\right)^2 \int_0^{\infty} \frac{dt}{\Delta^2} \sum_{s,s'} \left[2
m^2+(s s'-1)\Delta^2\right] \Theta\left( \tilde{\mu}_n-|{\cal
M}_s|\right) \left({\cal M}_s+{\cal M}_{s'} \right)
\nonumber \\&&
\Big\{\left({\cal M}_{s'}-{\cal M}_s \right)
\log\left(\frac{\tilde{\mu}_n+p_{F s}}{\tilde{\mu}_n-p_{F s}}
\right)+\eta \left({\cal M}_s+{\cal M}_{s'} \right)
\frac{p_0^2-\left({\cal M}_s-{\cal M}_{s'} \right)^2}{\Lambda}
\nonumber \\&&
 \Big[ 2 \Theta\left( 4 p_0^2 {\cal M}_s^2-(p_0^2+{\cal M}_s^2-{\cal
M}_{s'}^2)^2 \right) \arctan \left(\frac{p_0^2+{\cal M}_s^2-{\cal
M}_{s'}^2}{\eta \Lambda \tilde{\mu}_n} p_{F s}\right)-
\nonumber\\&& \Theta\left( (p_0^2+{\cal M}_s^2-{\cal
M}_{s'}^2)^2-4 p_0^2 {\cal M}_s^2 \right)\log \left(\frac{\Lambda
\tilde{\mu}_n+\eta(p_0^2+{\cal M}_s^2-{\cal M}_{s'}^2) p_{F
s})}{\Lambda \tilde{\mu}_n-\eta(p_0^2+{\cal M}_s^2-{\cal
M}_{s'}^2) p_{F s})} \right) \Big]\Big\}
 \label{AppB1}
\end{eqnarray}
where $\Delta=\sqrt{m^2+t}$, ${\cal M}_s=s \Delta-\kappa_n B$,
$p_{Fs}=\sqrt{\tilde{\mu}_n^2-{\cal M}_s^2}$,
$\eta=\text{sgn}(p_0)$, and $\Lambda=\sqrt{|4 p_0^2 {\cal
 M}_s^2-(p_0^2+{\cal M}_s^2-{\cal M}_{s'}^2)^2|}$.\\
  Although the
domain of integration is not bounded, the relation
\[\Theta\left(\tilde{\mu}_n-|{\cal M}_s|\right)\equiv
\Theta\left((\tilde{\mu}_n+s \kappa_n B)^2-m^2-t\right)\,
\Theta\left(\tilde{\mu}_n+s \kappa_n B-m\right) \] which is valid
for the conditions under consideration, establishes an upper limit
of
integration.\\
By performing the sum over $s'$ it can be rewritten as

\begin{eqnarray}
\text{Re} \, \Pi^{(n)}_0 (p_0)&=&\left(\frac{g_A}{\pi f_\pi}
\right)^2 \int_0^{\infty} \frac{dt}{2 \Delta^2} \sum_s \Theta\left(
\tilde{\mu}_n-|{\cal M}_s|\right)\left(A_{1 s}-\kappa_n B t A_{2 s}
\right)
 \label{AppB1} \\
 A_{1 s}&=&p_0 \, m^2 \frac{{\cal M}_s^2}{\lambda}\Bigg[ 2 \, \Theta\left(4 {\cal
 M}_s^2-p_0^2\right) \arctan\left( \frac{p_0 p_{F s}}{\tilde{\mu}_n \lambda} \right)+ \nonumber \\
 &&\Theta\left(p_0^2-4 {\cal M}_s^2\right) \log \left( \frac{\tilde{\mu}_n \lambda-p_0 p_{F s}}
 {\tilde{\mu}_n \lambda+p_0 p_{F s}} \right) \Bigg]
 \label{AppB2}\\
 A_{2 s}&=&s \Delta \log\left(\frac{\tilde{\mu}_n+p_{F s}}{\tilde{\mu}_n-p_{F s}}
 \right)+\eta \kappa_n B \frac{p_0^2-4 \Delta}{\lambda'}\Bigg[2 \, \Theta\left(4 p_0^2 {\cal
 M}_s^2-(p_0^2-4 s \kappa_n B \Delta)^2\right)\nonumber \\ &&
 \arctan\left(\eta p_{F s}\frac{p_0^2-s \kappa_n B \Delta}{\tilde{\mu}_n
 \lambda'}\right)+\Theta\left((p_0^2-4 s \kappa_n B \Delta)^2-4 p_0^2 {\cal
 M}_s^2\right)
 \nonumber \\ &&
 \log\left(\frac{\tilde{\mu}_n \lambda'-\eta (p_0^2-4 s \kappa_n B \Delta_2)p_{F s}}
 {\tilde{\mu}_n \lambda'+\eta (p_0^2-4 s \kappa_n B \Delta_2)p_{F s}} \right)
 \Bigg] \label{AppB3}
\end{eqnarray}
 $\lambda=\sqrt{|p_0^2-4 {\cal
 M}_s^2|}$, and $\lambda'=\sqrt{|4 p_0^2
{\cal
 M}_s^2-(p_0^2+4 s \kappa_n B \Delta_2)^2|}$.

\begin{eqnarray}
\text{Im} \, \Pi^{(n)}_0 (p_0)&=&\left(\frac{g_A}{4 f_\pi} \right)^2
\int_0^{\infty} \frac{dt}{\pi} \sum_{s,s'} \frac{s
s'}{\Delta^2}\left[ 2 m^2+(s s'-1)\Delta^2\right] \frac{({\cal
M}_s+{\cal M}_{s'})^2}{\Lambda_2}
\nonumber \\
&&\left[p_0^2-\left({\cal M}_s-{\cal M}_{s'}\right)^2 \right]\Bigg\{
\Theta\left(\tilde{\mu}_n-\frac{{\cal M}_s^2-{\cal
M}_{s'}^2+p_0^2}{2 p_0}\right) \Theta\left(\frac{(p_0-|{\cal
M}_s|)^2-{\cal M}_{s'}^2}{2 p_0}\right) \nonumber \\&& \left[2\,
\Theta\left(\tilde{\mu}_n-\frac{{\cal M}_s^2-{\cal
M}_{s'}^2-p_0^2}{2
p_0}\right)-1\right]+\Theta\left(\tilde{\mu}_n-\frac{{\cal
M}_s^2-{\cal M}_{s'}^2-p_0^2}{2 p_0}\right) \nonumber \\ &&
\Theta\left(\frac{{\cal M}_s^2-(p_0^2+|{\cal M}_{s'}|)^2}{2
p_0}\right)
 \Bigg\}
 \label{AppB}
\end{eqnarray}

\begin{eqnarray}
\text{Re} \, \Pi^{(p)}_0 (p_0)&=&\left(\frac{g_A}{\pi f_\pi}
\right)^2 q B \left[\Theta\left(\tilde{\mu}_p-|{\cal
M}_0|\right)A_0 +\frac{1}{16}\sum_{n, s, s'}
\Theta\left(\tilde{\mu}_p-|{\cal M}_{n s}|\right) A_{n s
s'}\right]
\nonumber \\
A_0&=&p_0 \frac{{\cal M}_0^2}{\lambda_0}\Bigg[ 2 \,
\Theta\left(4{\cal
 M}_0^2-p_0^2\right) \arctan\left( \frac{p_0 p_{F 0}}{\tilde{\mu}_p \lambda_0} \right)+
\Theta\left(p_0^2-4 {\cal M}_0^2\right) \log \left( \frac{\tilde{\mu}_p \lambda_0-p_0 p_{F 0}}{\tilde{\mu}_p \lambda_0+p_0 p_{F 0}} \right) \Bigg]
\nonumber \\
A_{n s s'}&=&\frac{\Delta_n+s \,m}{\Delta_n}\frac{\Delta_n+s'
m}{\Delta_n}\left(1-2\,
\frac{m-s'\Delta_n}{m+s\Delta_n}+\frac{m-s\Delta_n}{m+s\Delta_n}\frac{m-s'\Delta_n}{m+s'\Delta_n}\right)
\nonumber\\
&&\Bigg\{\left({\cal M}_{n s'}^2-{\cal M}_{n
s'}^2\right)\log\left(\frac{\tilde{\mu}_p+p_{F n
s}}{\tilde{\mu}_p-p_{F n s}} \right)+\frac{\eta}{\Lambda_1} \left(
{\cal M}_{n s}+{\cal M}_{n s'}\right)^2 \left[p_0^2-\left( {\cal
M}_{n s}-{\cal M}_{n s'}\right)^2 \right]\nonumber \\ && \Bigg[2 \,
\Theta\left(4 p_0^2 {\cal
 M}_{n s}^2-(p_0^2+{\cal M}_{n s}^2-{\cal M}_{n s'}^2)^2\right)
 \arctan\left(\eta p_{F n s}\frac{p_0^2+{\cal M}_{n s}^2-{\cal M}_{n
 s'}^2}{\tilde{\mu}_p
 \Lambda_1}\right)
 \nonumber \\ &&
 +\Theta\left((p_0^2+{\cal M}_{n s}^2-{\cal M}_{n s'}^2)^2-4 p_0^2 {\cal
 M}_{n s}^2\right) \log\left(\frac{\tilde{\mu}_p \Lambda_1-\eta (p_0^2+{\cal M}_{n s}^2-{\cal M}_{n
 s'}^2)p_{F n s}}
 {\tilde{\mu}_p \Lambda_1+\eta (p_0^2+{\cal M}_{n s}^2-{\cal M}_{n
 s'}^2)p_{F n s}} \right)
 \Bigg]\Bigg\}\nonumber
\end{eqnarray}

with ${\cal M}_0=m-\kappa_p B$, ${\cal M}_{n s}=s \Delta_n-\kappa_p
B$, $p_{F 0}=\sqrt{\tilde{\mu}_p^2-{\cal M}_0^2}$, $p_{F n
s}=\sqrt{\tilde{\mu}_p^2-{\cal M}_{n s}^2}$,
$\lambda_0=\sqrt{|p_0^2-4 {\cal M}_0^2|}$, and $\Lambda_1=\sqrt{|4
p_0^2 {\cal M}_{n s}^2-(p_0^2+{\cal M}_{n s}^2-{\cal M}_{n
s'}^2)^2|}$.
\begin{eqnarray}
\text{Im}\,\Pi_0^{(p)}(p_0)&=&\left(\frac{g_A}{f_\pi} \right)^2
\frac{q B}{2 \pi}\left(C_0+\frac{1}{8} \sum_{n s s'} C_{n s
s'}\right)
\nonumber \\
C_0&=&2 \eta p_0 \frac{{\cal
M}_0^2}{\lambda_0}\Big\{\Theta(-p_0-2{\cal M}_0)
\Theta(\tilde{\mu}_p+p_0/2)+\Theta(p_0-2{\cal M}_0)
\Theta(\tilde{\mu}_p-p_0/2)\nonumber
\\ &&\left[1-2\,\Theta(\tilde{\mu}_p+p_0/2) \right] \Big\}
\nonumber \\
C_{n s s'}&=&(\Delta_n+s m)(\Delta_n+s' m)\left(\frac{{\cal M}_{n
s}-{\cal M}_{n s'}}{\Delta_n} \right)^2 \frac{p_0^2-({\cal M}_{n
s}-{\cal M}_{n s'})^2}{\Lambda_1}
\nonumber \\
&&\left(1-2\,
\frac{m-s'\Delta_n}{m+s\Delta_n}+\frac{m-s\Delta_n}{m+s\Delta_n}\frac{m-s'\Delta_n}{m+s'\Delta_n}\right)\Bigg\{\Theta\left(\frac{(p_0-{\cal
M}_{n s})^2-{\cal M}_{n s'}^2}{2 p_0}\right)
\nonumber \\
&& \Theta\left(\tilde{\mu}_p-\frac{p_0^2+{\cal M}_{n s}^2-{\cal
M}_{n s'}^2}{2 p_0}\right)\left[1-2 \Theta\left(\tilde{\mu}_p+
\frac{p_0^2+{\cal M}_{n s'}^2-{\cal M}_{n s}^2}{2 p_0}\right)
\right]+\nonumber \\
&&\Theta\left(\frac{{\cal M}_{n s}^2-(p_0+{\cal M}_{n s'})^2}{2
p_0}\right) \Theta\left(\tilde{\mu}_p+\frac{p_0^2+{\cal M}_{n
s'}^2-{\cal M}_{n s}^2}{2 p_0}\right)\Bigg\}\nonumber
\end{eqnarray}
\begin{eqnarray}
\text{Re} \, \Pi_+ (p_0)&=&\left(\frac{g_A}{4 \pi f_\pi} \right)^2
\int_0^{\infty} \frac{dt}{\Delta}\, e^{-t/q B} \sum_{s'}
\Bigg\{\left(\Delta+s' m \right)\Big[\Theta\left(\tilde{\mu}_p -
|{\cal M}_0|\right) F_p(p_{F 0},{\cal M}_0,\lambda_0)
\nonumber \\
&&+\Theta\left(\tilde{\mu}_n- |{\cal M}_{s'}|\right) F_n(p_{F
0},{\cal M}_0,\lambda_0) \Big]+\sum_{n s} (-1)^n \frac{\Delta_n+s
m}{2 \Delta_n}
\nonumber \\
&&\left[ \left(\Delta +s' m\right)L_n-\left(\Delta-s'
m\right)\frac{s \Delta_n-m}{s \Delta_n+m}L_{n-1}-s' (s \Delta_n-m)
\left(L_n - L_{n-1}\right)\right]
\nonumber \\
&& \Big[\Theta\left(\tilde{\mu}_p - |{\cal M}_{n s}|\right) F_p(p_{F
n s},{\cal M}_{n s},\Lambda_1)+\Theta\left(\tilde{\mu}_n- |{\cal
M}_{s'}|\right) F_n(p_{F s'},{\cal M}_{n s},\Lambda_1) \Big] \Bigg\}
\label{RePiPlus} \\
F_a(x,M,U)&=&-I_a \left[4 p_0 x +\left(M^2-{\cal M}_{s'}^2
\right)\log\left(\frac{\tilde{\mu}_a+x}{\tilde{\mu}_a-x} \right)
\right]+\frac{\eta}{U} \left( {\cal
M}_{s'}+M\right)^2\left[p_0^2-( {\cal M}_{s'}-M)^2\right]
\nonumber \\
&&\Bigg\{ 2 \, \Theta\left(4 p_0^2 M^2-(p_0^2+M^2-{\cal
M}_{s'}^2)^2\right) \Bigg[ \arctan\left(\eta x
\frac{p_0^2+I_a\left(M^2-{\cal M}_{s'}^2\right)}{\tilde{\mu}_a
U}\right)
\nonumber \\
&&+I_a \arctan \left(2 \eta p_0 \frac{x}{U} \right)\Bigg]
+\Theta\left((p_0^2+M^2-{\cal M}_{s'}^2)^2-4 p_0^2 M^2\right) I_a
\nonumber \\&&
 \Big[\log\left(\frac{\tilde{\mu}_a U-\eta  (p_0^2\, I_a+M^2-{\cal M}_{s'}^2)x}
 {\tilde{\mu}_a U+\eta (p_0^2 \,I_a+M^2-{\cal M}_{s'}^2)x} \right)+  \log\left(\frac{U-2 \eta p_0 x}
 {U+ 2 \eta p_0 x} \right)\Big] \Bigg\}\nonumber \\
\end{eqnarray}
where we have introduced the isospin projection number $I_p=1, \,
I_n=-1$, and all Laguerre functions have the same argument $L_k(2
t/qB)$.

\begin{eqnarray}
\text{Im} \, \Pi_+ (p_0)&=&\left(\frac{g_A}{4 f_\pi} \right)^2
\int_0^{\infty} \frac{dt}{\pi}\, \frac{e^{-t/q B}}{\Delta}
\sum_{s'}\Bigg\{2 \left(\Delta+ s'm \right) G({\cal
M}_0,\lambda_0)+\sum_{n s} (-1)^n \frac{\Delta_n+s m}{\Delta_n}
\nonumber \\
&& \left[ \left(\Delta+s' m \right)L_n-\left(\Delta-s'm
\right)\frac{s \Delta_n-m}{s \Delta_n+m}L_{n-1}-s' (s
\Delta_n-m)\left(L_n - L_{n-1}\right)\right] G({\cal M}_{n
s},\Lambda_1)  \Bigg\}
\nonumber \\
G(M,U)&=& \left( {\cal M}_{s'}+M\right)^2 \frac{p_0^2-( {\cal
M}_{s'}-M)^2}{U}\Bigg\{ \Theta\left(\frac{M^2-(p_0+|{\cal
M}_{s'}|)^2}{2 p_0} \right)
\Theta\left(\tilde{\mu}_n-\frac{M^2-{\cal M}_{s'}^2-p_0^2}{2 p_0}
\right)
\nonumber \\
&+& \Theta\left(\frac{(p_0-|M|)^2-{\cal M}_{s'}^2}{2 p_0}\right)
\Theta\left(\tilde{\mu}_p-\frac{M^2-{\cal M}_{s'}^2+p_0^2}{2 p_0}
\right)\Bigg[1-2  \Theta\left(\tilde{\mu}_n-\frac{M^2-{\cal
M}_{s'}^2-p_0^2}{2 p_0} \right)\Bigg] \Bigg\} \nonumber
\end{eqnarray}

\begin{eqnarray}
\text{Re} \, \Pi_- (p_0)&=&\left(\frac{g_A}{4 \pi f_\pi} \right)^2
\int_0^{\infty} \frac{dt}{\Delta}\, e^{-t/q B} \sum_{s'}
\Bigg\{\left(\Delta+s'm \right)\Big[\Theta\left(\tilde{\mu}_p -
|{\cal M}_0|\right) G_p(p_{F 0},{\cal M}_0,\lambda_0)
\nonumber \\
&&+\Theta\left(\tilde{\mu}_n- |{\cal M}_{s'}|\right) G_n(p_{F
0},{\cal M}_0,\lambda_0) \Big]+\sum_{n s} (-1)^n \frac{\Delta_n+s
m}{2 \Delta_n}
\nonumber \\
&& \left[ \left(\Delta +s' m\right)L_n-\left(\Delta-s'
m\right)\frac{s \Delta_n-m}{s \Delta_n+m}L_{n-1}-s' (s \Delta_n-m)
\left(L_n - L_{n-1}\right)\right]
\nonumber \\
&& \Big[\Theta\left(\tilde{\mu}_p - |{\cal M}_{n s}|\right) G_p(p_{F
n s},{\cal M}_{n s},\Lambda_1)+\Theta\left(\tilde{\mu}_n- |{\cal
M}_{s'}|\right) G_n(p_{F s'},{\cal M}_{n s},\Lambda_1) \Big] \Bigg\}
\label{RePiMinus} \\
G_a(x,M,U)&=&I_a \left[4 p_0 x +\left({\cal M}_{s'}^2-M^2
\right)\log\left(\frac{\tilde{\mu}_a+x}{\tilde{\mu}_a-x} \right)
\right]+\frac{\eta}{U} \left( {\cal
M}_{s'}+M\right)^2\left[p_0^2-( {\cal M}_{s'}-M)^2\right]
\nonumber \\
&&\Bigg\{ 2 \, \Theta\left(4 p_0^2 M^2-(p_0^2+M^2-{\cal
M}_{s'}^2)^2\right) \Bigg[ \arctan\left(\eta x
\frac{p_0^2+I_a\left(M^2-{\cal M}_{s'}^2\right)}{\tilde{\mu}_a
U}\right)
\nonumber \\
&&-I_a \arctan \left(2 \eta p_0 \frac{x}{U} \right)\Bigg]
+\Theta\left((p_0^2+M^2-{\cal M}_{s'}^2)^2-4 p_0^2 M^2\right) I_a
\nonumber \\&&
 \Bigg[ \log\left(\frac{\tilde{\mu}_a U-\eta  (p_0^2\, I_a+M^2-{\cal M}_{s'}^2)x}
 {\tilde{\mu}_a U+\eta (p_0^2 \,I_a+M^2-{\cal M}_{s'}^2)x} \right)+  \log\left(\frac{U+2 \eta p_0 x}
 {U- 2 \eta p_0 x} \right)\Bigg] \Bigg\}\nonumber
\end{eqnarray}

\begin{eqnarray}
\text{Im} \, \Pi_- (p_0)&=&\left(\frac{g_A}{4 f_\pi} \right)^2
\int_0^{\infty} \frac{dt}{\pi}\, \frac{e^{-t/q B}}{\Delta}
\sum_{s'}\Bigg\{2 \left(\Delta+s'm \right) H({\cal
M}_0,\lambda_0)+\sum_{n s} (-1)^n \frac{\Delta_n+s m}{\Delta_n}
\nonumber \\
&& \left[ \left(\Delta+s'm \right)L_n-\left(\Delta-s' m
\right)\frac{s \Delta_n-m}{s \Delta_n+m}L_{n-1}-s' (s
\Delta_n-m)\left(L_n -L_{n-1} \right)\right] H({\cal M}_{n
s},\Lambda_1) \Big] \Bigg\}
\nonumber \\
H(M,U)&=& \left( {\cal M}_{s'}+M\right)^2 \frac{p_0^2-( {\cal
M}_{s'}-M)^2}{U}\Bigg\{ \Theta\left(\frac{{\cal
M}_{s'}^2-(p_0+|M|)^2}{2 p_0} \right)
\Theta\left(\tilde{\mu}_p-\frac{{\cal M}_{s'}^2-M^2-p_0^2}{2 p_0}
\right)
\nonumber \\
&& +\Theta\left(\frac{(p_0-|{\cal M}_{s'}|)^2-M^2}{2 p_0} \right)
\Theta\left(\tilde{\mu}_n-\frac{{\cal M}_{s'}^2-M^2+p_0^2}{2 p_0}
\right)\nonumber \\
&&\Bigg[1-2 \Theta\left(\tilde{\mu}_p-\frac{{\cal
M}_{s'}^2-M^2-p_0^2}{2 p_0} \right)\Bigg] \Bigg\} \nonumber
\end{eqnarray}

The expressions for the pion polarizations $\Pi^{OPV}_-$ and
$\Pi^{OPV}_+$ are formally related by the simple transformation
$p_0 \rightarrow-p_0$.


\newpage
\begin{figure}
\includegraphics[width=0.8\textwidth]{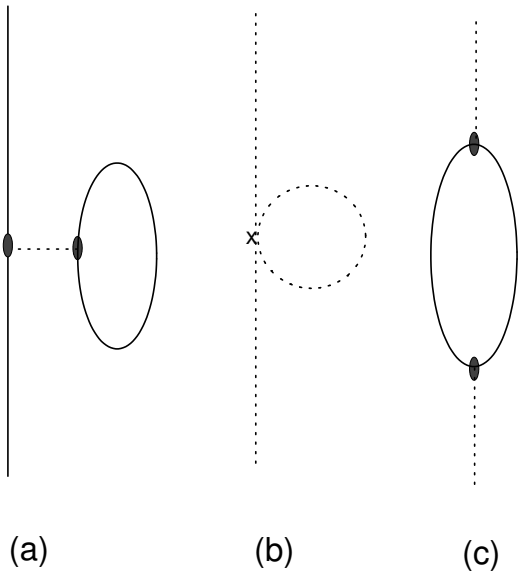}
\caption{\footnotesize The diagrams included in our Dyson-Schwinger
calculations. Solid lines represent fermion propagators, while
dashed lines represent the meson propagators. (a)The tadpole diagram
contributing to the mean field approach to the nucleon self-energy.
(b) The Weinberg-Tomozawa contribution to the pion polarization. (c)
The one pion exchange diagram for the pion polarization.}
\end{figure}

\newpage
\begin{figure}
\includegraphics[width=0.8\textwidth]{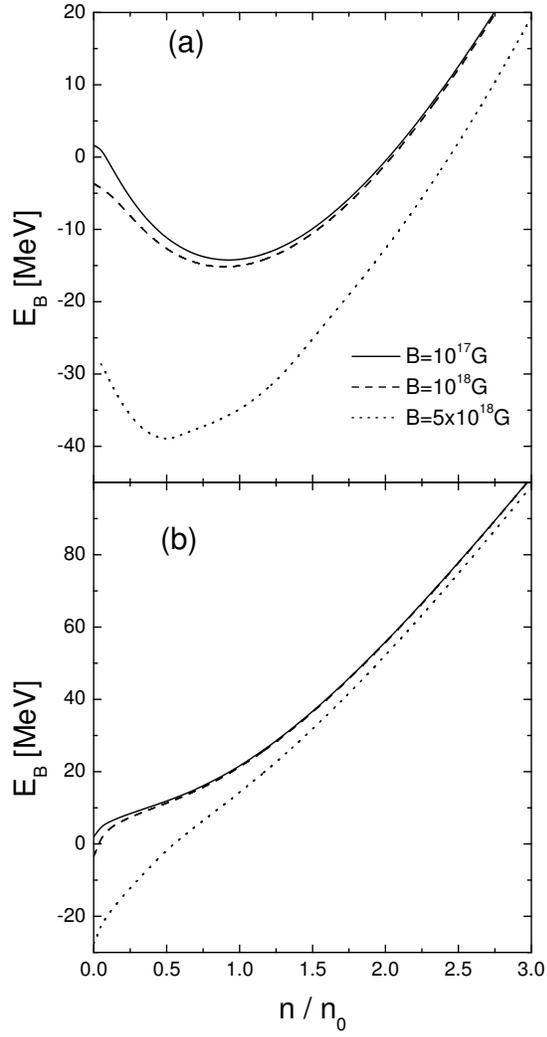}
\caption{\footnotesize The energy per particle (with the rest mass
subtracted) for (a) symmetric nuclear matter, and (b) pure neutron
matter, as a function of the density for different magnetic
intensities.}
\end{figure}
\newpage

\begin{figure}
\includegraphics[width=0.8\textwidth]{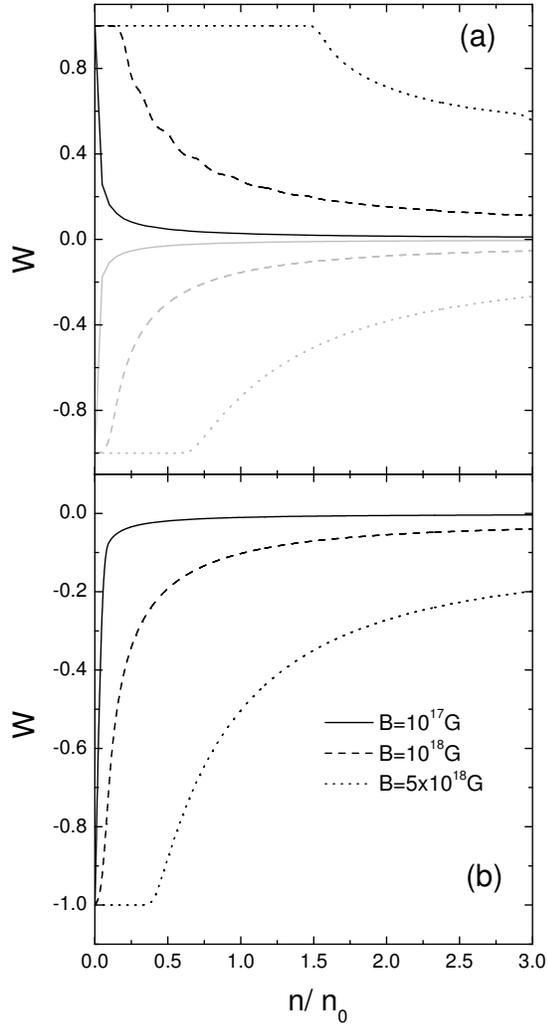}
\caption{\footnotesize The degree of polarization in (a) symmetric
nuclear matter, and (b) pure neutron matter as a function of the
density for different magnetic intensities. In the upper panel,
black lines correspond to the proton component and gray lines
correspond to the neutron case.}
\end{figure}

\begin{figure}
\includegraphics[width=0.8\textwidth]{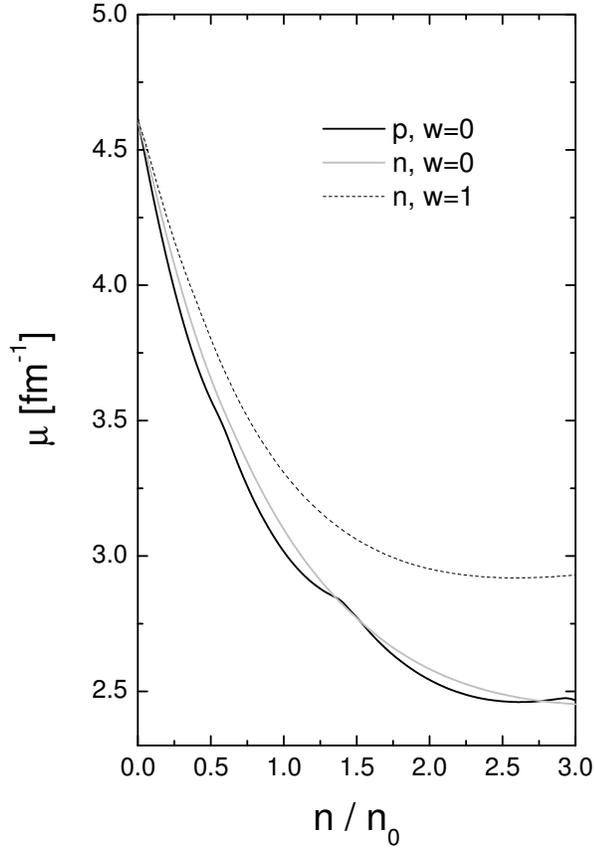}
\caption{\footnotesize The effective chemical potential as a
function of the density for $B=5 \times 10^{18}$ G. Here
$w=(n^{n}_B-n^{p}_B)/(n^{n}_B+n^{p}_B)$ is used. Solid lines
correspond to symmetric nuclear matter, and dashed lines correspond
to pure neutron matter. In the first case, a black line is used for
protons, and a gray line is used for neutrons.}
\end{figure}

\begin{figure}
\includegraphics[width=0.8\textwidth]{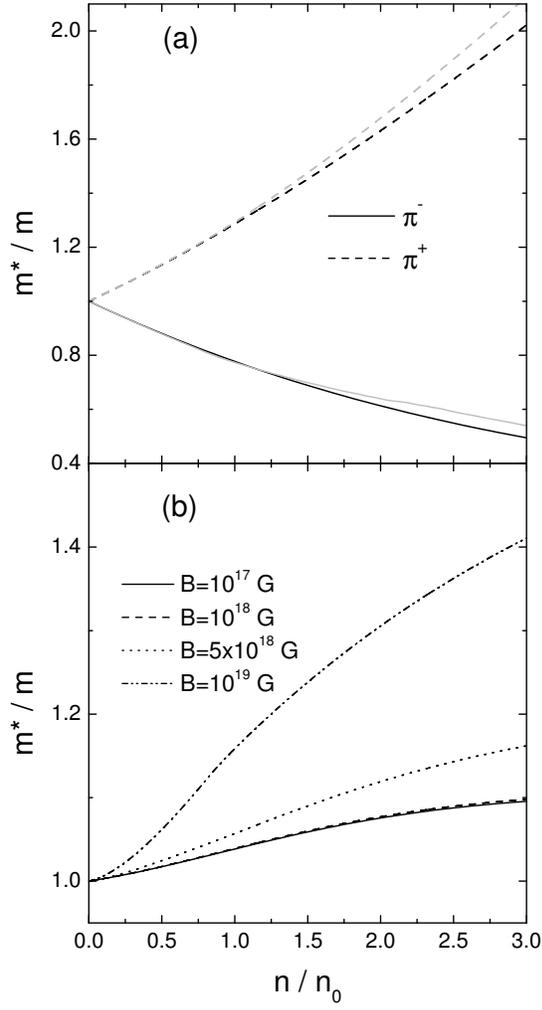}
\caption{\footnotesize The effective mass for (a) charged pions, and
 (b) the neutral pion as a function of the density for several magnetic
intensities in pure neutron matter.}
\end{figure}

\begin{figure}
\includegraphics[width=0.8\textwidth]{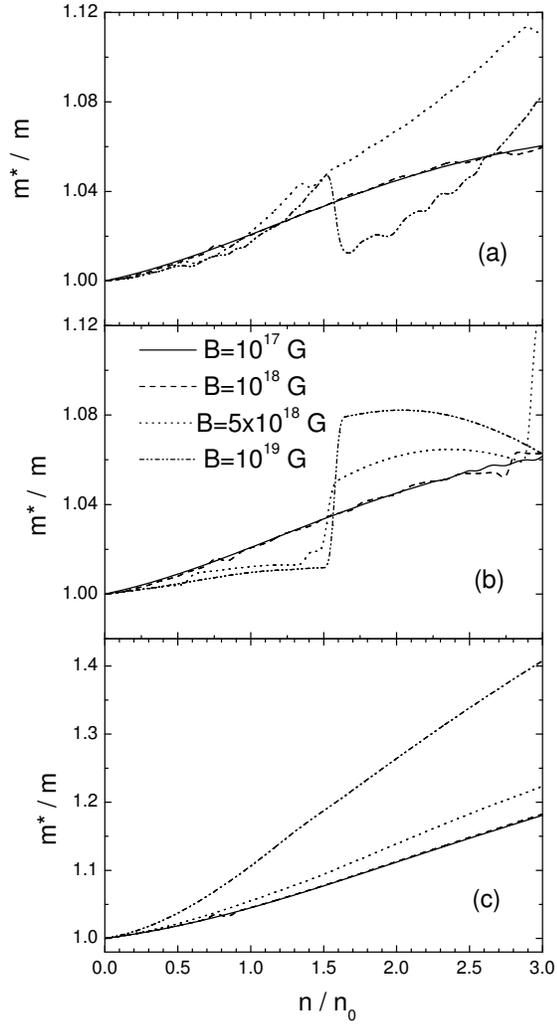}
\caption{\footnotesize The effective mass of (a) the negatively
charged pion, (b) the positively charged pion, and (c) the neutral
pion as a function of the density for several magnetic intensities
and isospin symmetric nuclear matter. }
\end{figure}

\begin{figure}
\includegraphics[width=0.8\textwidth]{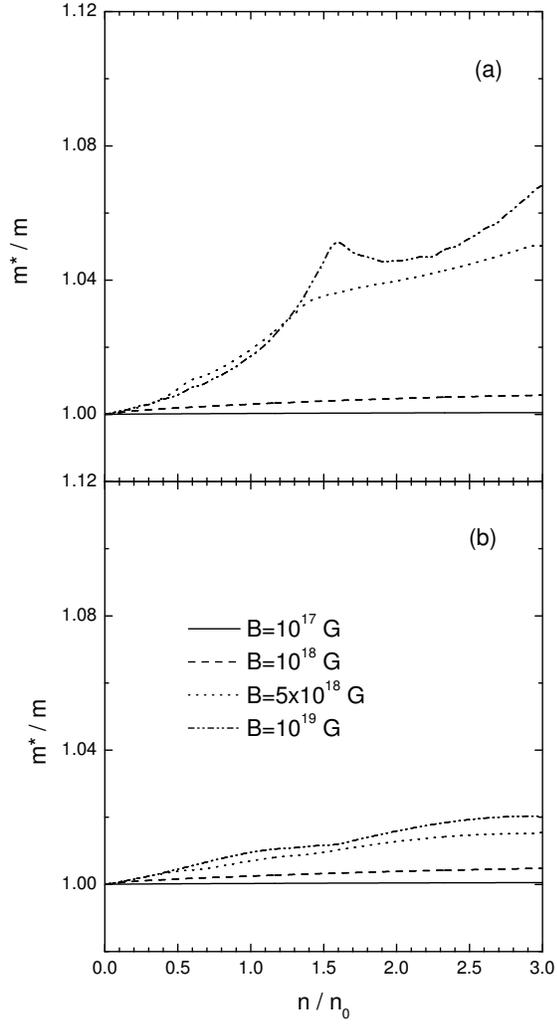}
\caption{\footnotesize The effective mass of the charged pions in
the lowest Landau level approximation for symmetric nuclear matter
as a function of the density. Results for (a)the negative pion  and
 (b)the positive pion are shown for several magnetic intensities.}
\end{figure}

\end{document}